\documentclass[11pt, a4paper]{article}
\usepackage{jheppub}
\usepackage{graphicx, bm, color, amssymb, amsmath,slashed,xcolor}
\usepackage{multirow}
\usepackage{float}
\usepackage[caption=false]{subfig}
\usepackage{tabularx}
\usepackage{hyperref}
\usepackage[normalem]{ulem}

\newcolumntype{L}[1]{>{\raggedright\arraybackslash}p{#1}}
\newcolumntype{C}[1]{>{\centering\arraybackslash}p{#1}}
\newcolumntype{R}[1]{>{\raggedleft\arraybackslash}p{#1}}

\begin{document}
\preprint{LAPTH-012/16, IPPP/16/22, DCPT/16/44}

\title{The Lepton Flavour Violating Higgs Decays at the HL-LHC and the ILC}
\author{Shankha Banerjee$^1$, Biplob Bhattacherjee$^2$, Manimala Mitra$^3$, Michael Spannowsky$^4$}
\affiliation[1]{LAPTH, Univ. de Savoie, CNRS, B.P.110, F-74941 Annecy-le-Vieux, France}
\affiliation[2]{Centre for High Energy Physics, Indian Institute of Science, 560012 Bangalore, INDIA }
\affiliation[3]{Indian Institute of Science Education and Research Mohali, Knowledge city, Sector 81, SAS Nagar, Manauli PO 140306.}
\affiliation[4]{Institute for Particle Physics Phenomenology, 
  Durham University, Durham DH1 3LE, United Kingdom.}
\abstract{Run-I results from the CMS collaboration show an excess of events in the decay $h \to \mu \tau_e$ with
a local significances of 2.4$\sigma$. This could be the first hint of flavour violation in the
Higgs sector. We summarise the bounds on the flavour violating Yukawa couplings from direct searches, low energy measurements 
and projected future experiments. We discuss the sensitivity of upcoming HL-LHC runs and future lepton colliders in measuring 
lepton-flavour violating couplings using an effective field theory framework. For the HL-LHC we find limits on 
BR($h\to \mu \tau$) and BR$(h\to e \tau) \lesssim \mathcal{O}(0.5)\%$ and on BR($h \to e \mu) \lesssim \mathcal{O}(0.02)\%$.
For an ILC with center-of-mass energy of 1 TeV we expect BR($h \to e \tau$) and BR($h\to \mu \tau$) to be measurable down to $\mathcal{O}(0.2)\%$.}

\maketitle
\section{Introduction}
\label{sec:1}

The Large Hadron Collider (LHC) has successfully discovered a scalar resonance of mass around 125 GeV~\cite{Aad:2012tfa, Chatrchyan:2012xdj}, 
with properties in close agreement with the Standard Model (SM) Higgs boson. Already now, both ATLAS and CMS collaborations have established its couplings to massive gauge bosons and photons to a high degree of precision and found no 
significant deviations from SM predictions \cite{ATLAS-CMS,Aad:2015gba,Khachatryan:2014jba}.

Direct limits on rather complex Higgs-fermion interactions were instead much less probed during initial LHC runs. Only fairly weak limits were obtained on Higgs couplings to $b$-quarks~\cite{ATLAS-CMS}, $\tau$-leptons \cite{Aad:2015vsa,Chatrchyan:2014nva} and $t$-quarks \cite{Khachatryan:2014qaa,CMS-ttbar,Aad:2014lma,Aad:2015iha,Aad:2015gra}.  Albeit the good overall agreement between theory and experiment, i.e.  the total signal strength measured from production and decay modes is $\mu=1.09 \pm 0.11$ \cite{ATLAS-CMS}, more conclusive evidence is required to establish that the observed scalar resonance is indeed  the SM Higgs boson. 

In particular the Higgs decays to the first and second generation fermions are yet to be observed.  
While searches have been performed by ATLAS and CMS to measure the Higgs decays to a $\mu^+\mu^-$ pair ~\cite{Aad:2014xva,Khachatryan:2014aep} and by CMS to an  $e^+e^-$ pair ~\cite{Khachatryan:2014aep}, only upper limits have been obtained so far. Indirect limits on these decays could be potentially obtained from a very precise measurement of the total Higgs decay width. While future linear colliders can play an important role in this~\cite{Peskin:2013xra,Peskin:2012we}, at present, the Higgs width is only fairly loosely bounded to $\Gamma_H < 22$ MeV (22.7 MeV) at 95$\%$ C.L from CMS (ATLAS)~\cite{Khachatryan:2014iha, Aad:2015xua} using highly model-dependent off-shell coupling measurements~\cite{Englert:2014ffa,Englert:2014aca}. Further, a global coupling fit to Higgs data 
indicates that the Higgs boson can have a sizable non-standard branching fraction, {\textit i.e.} BR$_{\textrm{non-std}} < 0.26$ at 95$\%$ C.L \cite{Giardino:2013bma}. Hence, within the present experimental findings exotic Higgs decays into first and second-generation fermions, gluons or missing energy can be significantly enhanced compared to SM predictions.

Among the many different plausible non-standard decay modes of the Higgs boson, one of the most intriguing are flavour violating Higgs decays. 
In the SM, these decays are highly suppressed, thus any experimental confirmation of such a process will be conclusive evidence of physics beyond the standard
model (BSM). 
During run-I of the LHC, a number of searches have been carried out both by CMS~\cite{Khachatryan:2015kon, CMS:2015udp} and 
 ATLAS~\cite{Moriond-2016,Aad:2015gha}. The result reported by CMS constrains the branching ratio BR$(h \to \mu \tau) < 1.51 $ $\%$ at 95$\%$ C.L, while the upper limit on the branching ratio reported by ATLAS is 1.43$\%$. 
Remarkably, both CMS and ATLAS reported a mild excess in the $h \to  \mu \tau_e$ channel with a local significances of $2.4 \sigma$ and $1.3 \sigma$  respective\footnote{ATLAS has studied 2 signal regions
 and the local significance in one of the signal regions (SR2) is about $2.2 \sigma$.}. These results can be explained with the best-fit branching ratio of BR$(h \to \tau \mu)=0.84\% (0.77\%)$ for CMS (ATLAS). In addition, CMS also looked  for flavour violation in the $e\mu$ and $e \tau$ channels~\cite{CMS:2015udp}. Recently, ATLAS updated their results from the $h \to e \tau$ analysis at 8 TeV~\cite{Moriond-2016}.

Apart from direct searches, flavour violating interactions of the Higgs boson can also be measured in low-energy observables, e.g. $\mu \to e \gamma$, $\tau \to \mu \gamma$, $\tau \to e \gamma$,  $\tau \to 3e$, $\mu \to 3e$ and $\mu-e$ conversion in nuclei. For a detailed discussion, see Refs.~\cite{Harnik:2012pb, Blankenburg:2012ex}. Hence, non-observation of these processes puts additional constraints on flavour violating couplings~\cite{Blankenburg:2012ex, Harnik:2012pb, Belusca-Maito:2016axk}. 

Assuming the low energy dynamics can be described in terms of an effective field theory (EFT), flavour violation in the Higgs sector is highly correlated with flavour violation in low energy processes \cite{Blankenburg:2012ex,Harnik:2012pb,Belusca-Maito:2016axk,Dorsner:2015mja}. 
However, in the presence of light degrees of freedom, this correlation might not hold. There have been several attempts to 
construct concrete models that can explain large branching ratios of the Higgs into non-degenerate fermion flavours, while 
simultaneously satisfying all low energy constraints. Lepton flavour violating Higgs decays have been discussed in the context of 
supersymmetry~\cite{Arhrib:2012ax, Abada:2014kba, Arganda:2015naa,Arganda:2015uca, Alvarado:2016par},
extended Higgs sectors~\cite{Das:2015zwa, Arroyo:2013tna, Kopp:2014rva,Sierra:2014nqa,Crivellin:2015mga,deLima:2015pqa,Das:2015kea, Mao:2015hwa,
Botella:2015hoa, Benbrik:2015evd,Omura:2015xcg,Zhang:2015csm, 
Bizot:2015qqo, Buschmann:2016uzg, Sher:2016rhh, Han:2016bvl} and other BSM models~\cite{Crivellin:2013hpa,Crivellin:2014cta,Dery:2014kxa,Campos:2014zaa,Heeck:2014qea,He:2015rqa, 
Cheung:2015yga,Feruglio:2015gka,Delepine:2015mca,Kosnik:2015lka,Baek:2015mea,Baek:2015fma, Hue:2015fbb, Chang:2016ave}. Some collider aspects in the
flavour violating sector have been studied in Refs.~\cite{Davidson:2012ds,Curtin:2013fra,Bressler:2014jta,Yue:2015dia,Bhattacherjee:2015sia}.

Thus, motivated by the recent searches carried out by CMS and ATLAS, we present a detailed analysis of flavour violation in direct and 
indirect experimental searches and compare their sensitivities. We first review the different experimental constraints on Higgs lepton-flavour violating and non-violating couplings in section~\ref{sec:2}. Here we also discuss some of the future limits of low energy experiments, e.g. MEG-II, Belle-II and super KEKB. After establishing the reach of low energy constraints using an EFT framework, we study the sensitivity of future high luminosity LHC runs on lepton flavour violating decays in section~\ref{sec:3}. We provide a 
case study for $h\to e \tau$ at the ILC in section~\ref{sec:4}. In sections~\ref{sec:3} and~\ref{sec:4}, we consider the 
respective LFV branching ratios as free parameters and evaluate how well they can be constrained in collider measurements. Finally we discuss our findings and summarise them in section~\ref{sec:5}.

\section{Higgs couplings in the lepton sector}
\label{sec:2}

In the SM, the Yukawa couplings are proportional to the masses of the fermions. Confirmation of this hypothesis requires independent measurements of the fermion masses and their 
coupling strength to the Higgs boson. That is a strenuous task, particularly for the Higgs boson couplings to the first and second generation fermions. A precise statement about the 
relation between Yukawa couplings and fermion masses is still at stake. Because of the observation of the recent excess reported
by CMS in the $h \to \mu \tau$ channel, we are restricting ourselves to a study of LFV interactions\footnote{A detailed study on the flavour
violating Higgs in the quark sector has been discussed, for example in Ref.~\cite{Harnik:2012pb}.}. In this section we  
review and update the existing searches on such LFV decays and summarise their bounds.

To give an interpretation of the measurements in terms of LFV interactions we consider an effective theory where the interaction between Higgs and fermions are given by the Yukawa interactions
\begin{equation}
 \mathcal{L}_Y = -m_i \bar{f_L^i}f_R^i - Y_{ij} (\bar{f_L^i}f_R^j)h +\; h.c..
 \label{eq:yuk}
\end{equation}
where we use $Y_{ii} = \frac{m_i}{v}$ and $f_L$ and $f_R$ are the charged leptons. The Yukawa coupling matrix is parametrised~\cite{Harnik:2012pb} by 
\begin{equation}
 Y_{ij} = \frac{m_i}{v}\delta_{ij} + \frac{v^2}{\sqrt{2} \Lambda^2} \hat{\lambda}_{ij},
\end{equation}
with $\hat{\lambda} = V_L \lambda^{'} V_R$. $V_L$ and $V_R$ are unitary matrices which diagonalise the mass matrix after
electroweak symmetry breaking (EWSB) and $-\frac{\lambda_{ij}^{'}}{\Lambda^2}$ are the coefficients of the gauge invariant 
dimension-6 operators,
\begin{equation}
 \Delta \mathcal{L}_Y = -\frac{\lambda_{ij}^{'}}{\Lambda^2} \bar{F_L^i}F_R^j H (H^{\dagger} H) + \; h.c,
\label{lfvyu1} 
\end{equation}
where $F_L$ is the fermion  doublet, $F_R$ is the singlet and $H$ is the SM scalar doublet.
For $\Lambda \to \infty$, we recover the SM Yukawa structure. There are also some gauge invariant dimension-6 operators involving
derivatives, induced by~\cite{Harnik:2012pb}
\begin{equation}
 \Delta \mathcal{L}_D = \frac{\lambda^{ij}_L}{\Lambda^2}(\bar{F_L^i} \gamma^{\mu} F_L^j)(H^{\dagger} i \overleftrightarrow{D_{\mu}} H) 
                      + \frac{\lambda^{ij}_R}{\Lambda^2}(\bar{F_R^i} \gamma^{\mu} F_R^j)(H^{\dagger} i \overleftrightarrow{D_{\mu}} H), 
\end{equation}
with $H^{\dagger} i \overleftrightarrow{D_{\mu}} H = H^{\dagger} i D_{\mu} H - (i D_{\mu} H^{\dagger})H$. However, these operators
do not contribute to the $Hf\bar{f}$ couplings as shown in Eq.~\ref{eq:yuk} after EWSB and hence we disregard further discussions of these operators in the
present study\footnote{In Ref.~\cite{Belusca-Maito:2016axk}, LFV has also been studied in the context of dipole operators of the form $H\bar{f}\sigma^{\mu\nu}fV_{\mu\nu}$. 
Such operators can induce flavour violating three body decays of the Higgs boson. However, precision constraints are stringent and render the prospects of discovering these 
decays at collider experiments slim.}.

\subsection{Present status of Higgs couplings in the lepton sector}
\label{sec:leptbound}

We list the current constraints on the Higgs Yukawa couplings in the lepton sector. First, we briefly summarise the status of the 
flavour diagonal and off-diagonal leptonic decays from direct searches during run-I at the LHC. Next we discuss the low energy
constraints on the LFV Yukawas in an EFT framework.

\subsubsection{Limits from direct searches}
\label{sec:2.2.1}

 \begin{itemize}
  \item $h \to e^+ e^-$ : An upper limit on its branching ratio has been obtained by CMS \cite{Khachatryan:2014aep} of BR$(h \to e^+ e^-) < 0.19$\%, which is about $3.7 \times 10^5$ times that of the SM expectation.
  
  \item $h \to \mu^+ \mu^-$ : ATLAS and CMS obtain an upper limit on the branching ratio of BR$(h \to \mu^+ \mu^-) < 0.15$\% \cite{Aad:2014xva}  and 
  BR$(h \to \mu^+ \mu^-) < 0.16$\% \cite{Khachatryan:2014aep}  respectively.
    
  \item $h \to \tau^+ \tau^-$ : Both ATLAS \cite{Aad:2015vsa} and CMS \cite{Chatrchyan:2014nva} have measured the Higgs boson coupling to a
  pair of $\tau$ leptons. For $m_h = 125.36 \, \textrm{GeV} \, (125 \, \textrm{GeV})$, ATLAS (CMS) has measured a signal strength
  of $1.43^{+0.43}_{-0.37} \, (0.78 \pm 0.27)$ in this channel.
  
  \item $h \to e \mu$ : CMS \cite{CMS:2015udp} sets a limit on BR$(h \to e \mu) < 0.036$\% at 95\% CL. One thus obtains
  
  \begin{equation}
   \sqrt{|Y_{e \mu}|^2 + |Y_{\mu e}|^2} < 5.43 \times 10^{-4}.
  \end{equation}
  

  \item $h \to e \tau$ : CMS \cite{CMS:2015udp} studied the $h \to e \tau_{\mu}$ and $h \to e \tau_{had}$ channels and
  obtained better sensitivity than the current indirect limits. They find BR$(h \to e \tau) < 0.69$\% at 95\% CL. From this limit one deduces
  
  \begin{equation}
   \sqrt{|Y_{e \tau}|^2 + |Y_{\tau e}|^2} < 2.41 \times 10^{-3}.
  \end{equation}
On the other hand, ATLAS obtained a weaker limit BR$(h \to e \tau) < 1.04$\% at 95\% CL \cite{Moriond-2016}.  
  
  \item $h \to \mu \tau$ : The search was conducted in the channel $p p \to h \to \mu \tau$, followed by the leptonic as well 
  as hadronic decays of $\tau$. CMS reported a slight excess of events around $m_h = 125$ GeV in the $h \to \mu \tau_e$ channel 
  with a local significance of $2.4\sigma$~\cite{Khachatryan:2015kon}. From this they obtained 
  BR$(h \to \mu \tau) < 1.51\%$ at 95$\%$ CL with a best-fit of $(0.84^{+0.39}_{-0.37})\%$. ATLAS has set an upper limit of 
  1.43\% on this branching ratio at 95\% CL~\cite{Moriond-2016}. From the upper limit on BR$(h \to \mu \tau)$ from CMS one obtains

  \begin{equation}
   \sqrt{|Y_{\mu \tau}|^2 + |Y_{\tau \mu}|^2} < 3.6 \, \times 10^{-3}.
  \end{equation}

  In our calculation, we have used that the partial decay width of 125 GeV Higgs into two fermions is, 
  
  \begin{equation}
   \Gamma(h \to \ell_{\alpha} \ell_{\beta})=\frac{m_h}{8 \pi}( |Y_{\ell_{\alpha} \ell_{\beta}}|^2+ |Y_{\ell_{\beta} \ell_{\alpha}}|^2), 
  \end{equation}  
  where $\ell_{\alpha} = \ell_{\beta} = e,\mu,\tau$ and $\alpha \neq \beta$. The branching ratio for this decay mode is 
\begin{equation}
\rm{BR}(h\to \ell_{\alpha} \ell_{\beta})=\frac{\Gamma(h \to \ell_{\alpha} \ell_{\beta})}{\Gamma_{\rm{SM}}+\Gamma(h \to \ell_{\alpha} \ell_{\beta})},
\end{equation}
where $\Gamma_{\rm{SM}} = 4.1$ MeV.
\end{itemize} 
  
\subsubsection{Limits from low-energy measurements}
\label{sec:2.2.2}

We motivate below the different bounds on these Yukawa couplings that emerge from the low energy flavour violating processes
and summarise these limits in Table~\ref{tab:flvsum}. 

\begin{itemize}
\item
$Y_{\mu \tau}$ is constrained by the non-observation of processes like $\tau \to \mu \gamma$ and  $\tau \to 3 \mu$. 
The  branching ratio of the process $\tau \to \mu \gamma$ is bounded by 
BR$(\tau \to \mu \gamma) < 4.4 \times 10^{-8}$ at 90$\%$ C.L.~\cite{Beringer:1900zz,PDG}. Assuming the low energy dynamics to be 
governed by the two flavour violating couplings $Y_{\mu \tau}$ and $Y_{\tau \mu}$, the decay width of this process reads 
as~\cite{Harnik:2012pb}
\begin{equation}
\Gamma ( \tau \to \mu \gamma)=\frac{\alpha m^5_{\tau}}{64 \pi^4} (|c_L|^2+|c_R|^2), 
\label{lfvtautomugamma}
\end{equation}
where the Wilson coefficients, $c_L$ and $c_R$ at one loop are given by
\begin{equation}
c_{L(R)} \sim \frac{1}{3 m^2_h}Y_{\tau \tau} Y_{\tau \mu} \left (-1+ \frac{3}{4} \rm{log}\frac{{m}^2_h}{{m}^2_{\tau}} \right ).
\end{equation}
In the above expression, for simplicity, we assume that $Y_{\tau \mu} = Y_{\mu \tau}$. Higher-order corrections to
$c_{L/R}$ receive contributions from the top Yukawa coupling and hence can be large~\cite{Harnik:2012pb}. For example, including two-loop contributions, $c_{L/R}$ increases by a factor $\mathcal{O}({4})$. For a full discussion on the dependence of 
$Y_{tt}$ we refer to Ref.~\cite{Harnik:2012pb}. Hence by assuming $Y_{\tau \mu} = Y_{\mu \tau}$ and also a SM-like 
$Y_{\tau\tau}$, we obtain $Y_{\mu \tau} \lesssim 0.011$. 
 
In presence of flavour violating Yukawa interactions, the process $\tau \to 3 \mu$ has a decay width of
\begin{equation}
\Gamma ( \tau \to 3 \mu)=\frac{\alpha^2 m^5_{\tau}}{6(2 \pi)^5} (|c_L|^2+|c_R|^2).
\label{lfvtauto3}
\end{equation}
Taking into account the one-loop contribution, the constraint on the Yukawa coupling is,
$Y_{\mu \tau} < 0.177$. Similar to the previous limit, this  
also depends on the couplings $Y_{\tau \tau}, Y_{\mu \mu} \; \textrm{and } Y_{tt}$ and their SM values have been assumed
in deriving the limit on $Y_{\mu \tau}$. This limit is weaker than $\tau \to \mu \gamma$ due to an additional factor of 
$\alpha$ (where $\alpha$ is the fine-structure constant).

\item
The coupling $Y_{\tau e}$ is also constrained from similar low energy flavour violating processes, such as, 
$\tau \to e \gamma$ and $\tau \to 3 e$. The decay width for $\tau \to e \gamma$ has a similar expression as given in 
Eq.~\ref{lfvtautomugamma}, where $\mu $ has to be replaced by $e$~\cite{Harnik:2012pb}. Under similar assumptions, one obtains
$Y_{\tau e} \lesssim 0.01$ from $\tau \to e \gamma$, while from $\tau \to 3 e$, this is relaxed owing to the extra $\alpha$ 
factor. Assuming $Y_{\tau e}=Y_{e \tau}$, we show the bounds on the Yukawas in Table.~\ref{tab:flvsum}. 

\item
The Yukawa coupling $Y_{\mu e}$ is severely constrained by the $\mu \to e \gamma$ limit from MEG \cite{Adam:2013mnn}. The 
Wilson coefficients have similar expression as given in Eq.~\ref{lfvtautomugamma}, with $\{\tau,\mu\} \to \{\mu, e\}$. The two 
loop contribution proportional to the top Yukawa coupling can be large~\cite{Harnik:2012pb}. The updated limit on the branching ratio,
\textit{viz.}, BR$(\mu \to e \gamma) < 5.7 \times 10^{-13}$~\cite{Beringer:1900zz} imposes a tight constraint on the Yukawa 
$\sqrt{|Y_{\mu e}|^2+|Y_{e \mu}|^2} \lesssim 1.75 \times 10^{-6}$.  Assuming, 
$Y_{\mu e }= Y_{e \mu}$ this constrains $Y_{\mu e} \lesssim 1.24 \times 10^{-6}$. The other LFV process,\textit{i.e}, $\mu \to 3 e$ constrains
$Y_{\mu e} \lesssim 2.19 \times 10^{-5}$.

\item
In addition to the above constraints, for complex Yukawa couplings, severe constraints appear from the electric 
dipole moment measurement. The electric dipole moment for the electron is $|d_e| \le 10.5 \pm 0.07 \times 10^{-26}e$ cm~\cite{Beringer:1900zz},
which constrains the complex Yukawas $|\mathrm{Im}(Y_{e \tau }Y_{\tau e})| \lesssim 1.1 \times 10^{-8}$ and, 
$|\mathrm{Im}(Y_{e \mu }Y_{\mu e})| \lesssim 9.8 \times 10^{-8}$.

From the electric dipole measurement of muons, we have $-10 \times 10^{-20}e \textrm{ cm } < d_{\mu} < 8 \times 10^{-20}e \textrm{ cm}$~\cite{PDG2}.
This puts a weak constraint on $-0.8 \lesssim |\mathrm{Im}(Y_{\mu \tau }Y_{\tau \mu})| \lesssim 1.0$~\cite{Harnik:2012pb}.

\item
The stringent constraint  on $\mu \to e \gamma$ branching ratio can also be used to set the limit on the product of the 
flavour violating couplings $Y_{\tau \mu}$ and $Y_{\tau e}$. The limit from MEG, 
BR$(\mu \to e \gamma) < 5.7 \times 10^{-13}$~\cite{Beringer:1900zz}, imposes the constraint 
$(|Y_{\tau \mu} Y_{e \tau}|^2+|Y_{\mu \tau} Y_{\tau e}|^2)^{\frac{1}{4}} < 2.373 \times 10^{-4}$ at one loop level. Again assuming 
$Y_{\tau \mu}=Y_{\mu \tau}$ and $Y_{\tau e }=Y_{e \tau}$, 
one obtains, $Y_{\tau \mu } Y_{e \tau} \lesssim 3.98 \times 10^{-8}$. {{Note that, in the limit $Y_{\mu e}=0$, the two loop Barr-Zee diagram does not occur in this case.}}

\item
Besides, there are also constraints on the LFV Yukawa couplings from muonium-antimuonium 
oscillations~\cite{Willmann:1998gd,Clark:2003tv}, magnetic dipole moments~\cite{Nakamura:2010zzi,Bennett:2006fi} and from
$\mu \to e$ conversions in nuclei~\cite{Harnik:2012pb,vanderSchaaf:2003ti}. Furthermore, there are constraints from LEP which excludes
$\sqrt{|Y_{\ell e}|^2+|Y_{e \ell}|^2} < \textrm{ few } \times 10^{-1}$~\cite{Alcaraz:2006mx}. These constraints are 
tabulated in Table~\ref{tab:flvsum}. For a detailed review on these constraints, we refer the reader to Ref.~\cite{Harnik:2012pb}.

\end{itemize}
\begin{table}
\footnotesize
\centering
\begin{tabular}{|c|c|c|}
\hline
Searches  & Experimental limit on   & Limits on Yukawas  \\ 
 &  branching ratios & \\ \hline \hline
$\tau \to \mu \gamma$                           & $4.4 \times 10^{-8}$~\cite{Beringer:1900zz,PDG}  & $Y_{\mu \tau} < $ 0.011 \\ 
$\tau \to 3 \mu $                               & $2.1 \times 10^{-8}$~\cite{Beringer:1900zz,PDG}  & $Y_{\mu \tau}< $ 0.176 \\
Muon EDM                                        & $-10 \times 10^{-20}e \textrm{ cm } < $          & $-0.8 \lesssim$ \\
                                                & $|d_{\mu}| $                                     & $|\rm{Im}(Y_{\mu \tau }Y_{\tau \mu})| \lesssim 1.0$ \\
                                                & $< 8 \times 10^{-20}e \textrm{ cm}$~\cite{PDG2}  &  \\
Muon $g-2$                                      & $-$                                              & $\rm{Re}(Y_{\mu \tau} Y_{\tau \mu}) < (2.7\pm 0.75) \times 10^{-3}$ \\                                                   
$\tau \to \mu \gamma$ (f)                       & $ 10^{-9}$~\cite{Wang:2015kmm}                   & $Y_{\mu \tau} < $0.0017 \\ 
(Belle-II/super KEKB)                           &                                                  & \\
\hline \hline 
$\tau \to e \gamma$                             & $3.3 \times 10^{-8}$~\cite{Beringer:1900zz,PDG}  & $Y_{e \tau} < $0.0099 \\ 
$\tau \to 3 e$                                  & $2.7 \times 10^{-8}$~\cite{Beringer:1900zz,PDG}  & $Y_{e \tau}< $ 0.085 \\
Electron $g-2$                                  & $-$                                              & $\rm{Re}(Y_{e \tau} Y_{\tau e}) < [-2.1,2.9] \times 10^{-3}$ \\
Electron EDM                                    & $|d_e| \le 0.105 \times 10^{-26}$ e cm           & $|\rm{Im}(Y_{e\tau}Y_{\tau e})| < 1.1\times 10^{-8}$ \\
$\tau \to e \gamma$ (f)                         & $ 10^{-9}$~\cite{Wang:2015kmm}                   & $Y_{e \tau} < $0.00172 \\
(Belle-II/super KEKB)                           &                                                  & \\
\hline \hline
$\mu \to e \gamma$                              & $5.7 \times 10^{-13}$~\cite{Beringer:1900zz,PDG} & $Y_{\mu e} < $ 1.24 $\times 10^{-6}$ \\ 
$\mu \to 3 e$                                   & $1.0 \times 10^{-12}$~\cite{Beringer:1900zz,PDG} & $Y_{\mu e} < $ 2.19 $\times 10^{-5}$ \\ 
Electron $g-2$                                  & $-$                                              & $\rm{Re}(Y_{e \mu} Y_{\mu e}) < [-0.019,0.026]$ \\
Electron EDM                                    & $|d_e| \le 0.105 \times 10^{-26}$ e cm           & $|\rm{Im}(Y_{e\mu}Y_{\mu e})| < 9.8\times 10^{-8}$ \\
$\mu \to e$ conversion                          & $-$                                              & $Y_{\mu e} < $ 8.49 $\times 10^{-6}$ \\
$M-\bar{M}$ oscillations                        & $-$                                              & $|Y_{\mu e} + Y_{e \mu}^*| < 0.079$ \\
$\mu \to e \gamma$ (f) (MEG-II)                 & $4  \times 10^{-14}$~\cite{Renga:2014xra}        & $Y_{\mu e} < $3.28 $\times 10^{-7}$ \\ 
\hline \hline
$\mu \to e \gamma$                              & $5.7 \times 10^{-13}$                            & $Y_{\mu \tau} Y_{e \tau} < $3.98 $\times 10^{-8}$ \\ 
\hline \hline
$h \to \tau \mu$ (CMS)                          & 1.51$\%$~\cite{Khachatryan:2015kon}              & $Y_{\mu \tau} < 2.55 \times 10^{-3}$ \\
                                                & 0.84$\%$                                         & $Y_{\mu \tau} = 1.87 \times 10^{-3}$ \\

$h \to \tau \mu$ (ATLAS)                        & 1.43$\%$~\cite{Moriond-2016}                     & $Y_{\mu \tau} < 2.45 \times 10^{-3}$ \\
                                                                                              
                                                & 0.77$\%$~\cite{Aad:2015gha}                      & $Y_{\mu \tau} = 1.79 \times 10^{-3}$ \\

\hline \hline
$h \to \tau \mu$ (CMS)+ $\mu \to e \gamma$      & 0.84$\%$, $5.7 \times 10^{-13}$                  & $Y_{e \tau} < 2.13 \times 10^{-5}$ \\
$h \to \tau \mu$ (ATLAS)+ $\mu \to e \gamma$    & 0.77$\%$, $5.7 \times 10^{-13}$                  & $Y_{e \tau} < 2.23 \times 10^{-5}$ \\
\hline \hline
$h \to \tau e$ (CMS)                            & 0.69\%~\cite{CMS:2015udp}                        & $Y_{e \tau} < 1.69 \times 10^{-3}$ \\
$h \to \tau e$ (ATLAS)                          & 1.04\%~\cite{Moriond-2016}                       & $Y_{e \tau} < 2.08 \times 10^{-3}$ \\
\hline \hline
$h \to e \mu$ (CMS)                             & 3.6$\times 10^{-2}$\%~\cite{CMS:2015udp}         & $Y_{\mu e} < 3.85 \times 10^{-4}$ \\
\hline \hline
\end{tabular}
\normalsize
\caption{The low energy flavour violating  processes and upper limit on the Yukawa couplings. For simplicity, we assume the 
Yukawas $Y_{\alpha \beta}=Y_{\beta \alpha}$. The index $(f)$ refers to the prospective future measurements.}
\label{tab:flvsum}
 \end{table}
 
\subsubsection{Future limits}

In this subsection we briefly discuss the expected future limits on the flavour diagonal and flavour violating couplings
of the Higgs boson in the lepton sector.

\begin{itemize}
 \item $h \to e^+ e^-$ : Ref.~\cite{Altmannshofer:2015qra} discusses the present and future bounds on the $Y_{ee}$ Yukawa coupling. 
  The ACME bound on the EDM puts a strong constraint on the imaginary part of this Yukawa, i.e. $< 1.7 \times 10^{-2}$ 
  times the SM electron Yukawa, $Y_e$. The deviations of the real part are far less constrained. The authors find that the 
  constraint coming from 8 TeV LHC data is the strongest. A limit $\kappa_e < 611$ can be derived, where $\kappa_e$ is a multiplicative 
  complex parameter to the SM Yukawa, which parametrises the deviation from the SM Yukawa coupling. Noting that for the 14 TeV 
  run, the production cross-section of the Higgs will increase by a factor of $\sim 2.5$, they project that $\kappa_e \sim 260 (150)$ with 
  $\mathcal{L} = 300 (3000)\, \textrm{fb}^{-1}$. For a 100 TeV $pp$ collider with $\mathcal{L} = 3000 \, \textrm{fb}^{-1}$,
  $\kappa_e \sim 75$.
  
 \item $h \to \mu^+ \mu^-$ :   In Ref.~\cite{Han:2002gp}, it is mentioned that by combining the gluon fusion and 
  weak boson fusion channels, it is possible to obtain a 3$\sigma$ significance for $h \to \mu^+ \mu^-$ at an integrated 
  luminosity of $300$ fb$^{-1}$. It is also projected by CMS and ATLAS \cite{CMS-ATLAS-talk, ATL-PHYS-PUB-2013-014} that for 
  the 14 TeV run with an integrated luminosity of around 1200 fb$^{-1}$, one can observe the $h \to \mu^+ \mu^-$ mode with a 
  5$\sigma$ significance.
  
  \item $h \to \tau^+ \tau^-$ : Future runs of the LHC and the ILC are expected to improve the sensitivity of this coupling. 
  From Ref.~\cite{Peskin:2012we} one finds that the uncertainty on this coupling measurement decreases to about 12.5\% and about 1.5\%  
  respectively at future runs of LHC and ILC.
  
  \item $h \to e \mu$ : The improved sensitivity of MEG-II \cite{Renga:2014xra} will restrict $\sqrt{|Y_{\mu e}|^2+|Y_{e \mu}|^2} \le 4.64  \times 10^{-7}$.
  
  \item $h \to e \tau$ : The future sensitivity  of $\tau \to e \gamma$ with  BR$ \sim 10^{-9}$  will constrain the 
  Yukawas by a further order of magnitude $\sqrt{|Y_{\tau e}|^2+|Y_{e \tau}|^2} < 2.43 \times 10^{-3}$.
  
  \item $h \to \mu \tau$ : Future experiments such as, Belle-II/super (KEK) B factory with expected sensitivity 
  on BR$(\tau \to \mu \gamma) \sim 10^{-9}$ \cite{Wang:2015kmm}, will impose more stringent constraints on the flavour violating Yukawa 
  $\sqrt{|Y_{\tau \mu}|^2+|Y_{\mu \tau}|^2} < 2.41 \times 10^{-3}$. 
  
  \item Finally, we briefly mention the novel proposal given in Ref.~\cite{Delaunay:2016brc, Delaunay:2016zmu,Frugiuele:2016rii} which outlines an experimental technique to 
put bounds on the the flavour diagonal Higgs couplings to mostly the first generation fermions. The authors propose to achieve this 
by measuring isotope shifts in atomic clock transitions. This method can potentially bound the Higgs-light fermion couplings better than
the present and future runs of the LHC. By studying the isotope shift of the $Yb$ ion they show that one can bound the couplings to

\begin{equation}
 Y_u + 1.2 Y_d + 0.10 Y_s \lesssim 0.04 \Big( \frac{1.3 \times 10^{-3}}{Y_e}\Big) \Big(\frac{\Delta}{Hz}\Big),
\end{equation}
where $\Delta$ is the isotope shift measurement uncertainty.

\end{itemize}
\begin{figure}
\hspace*{-1.3cm}
\begin{center}
{
\subfloat[]{\includegraphics[width=7.8cm]{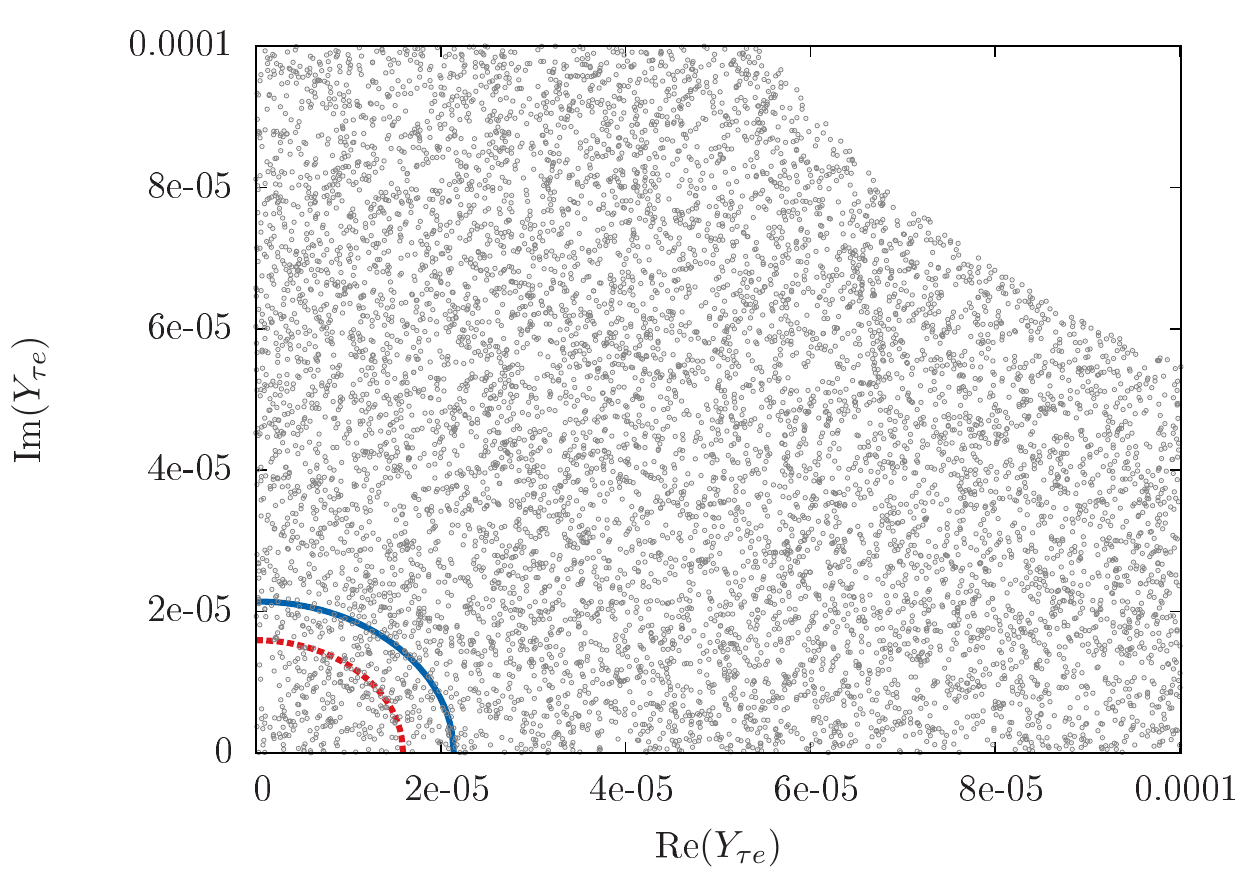}\label{}}
\subfloat[]{\includegraphics[width=7.8cm]{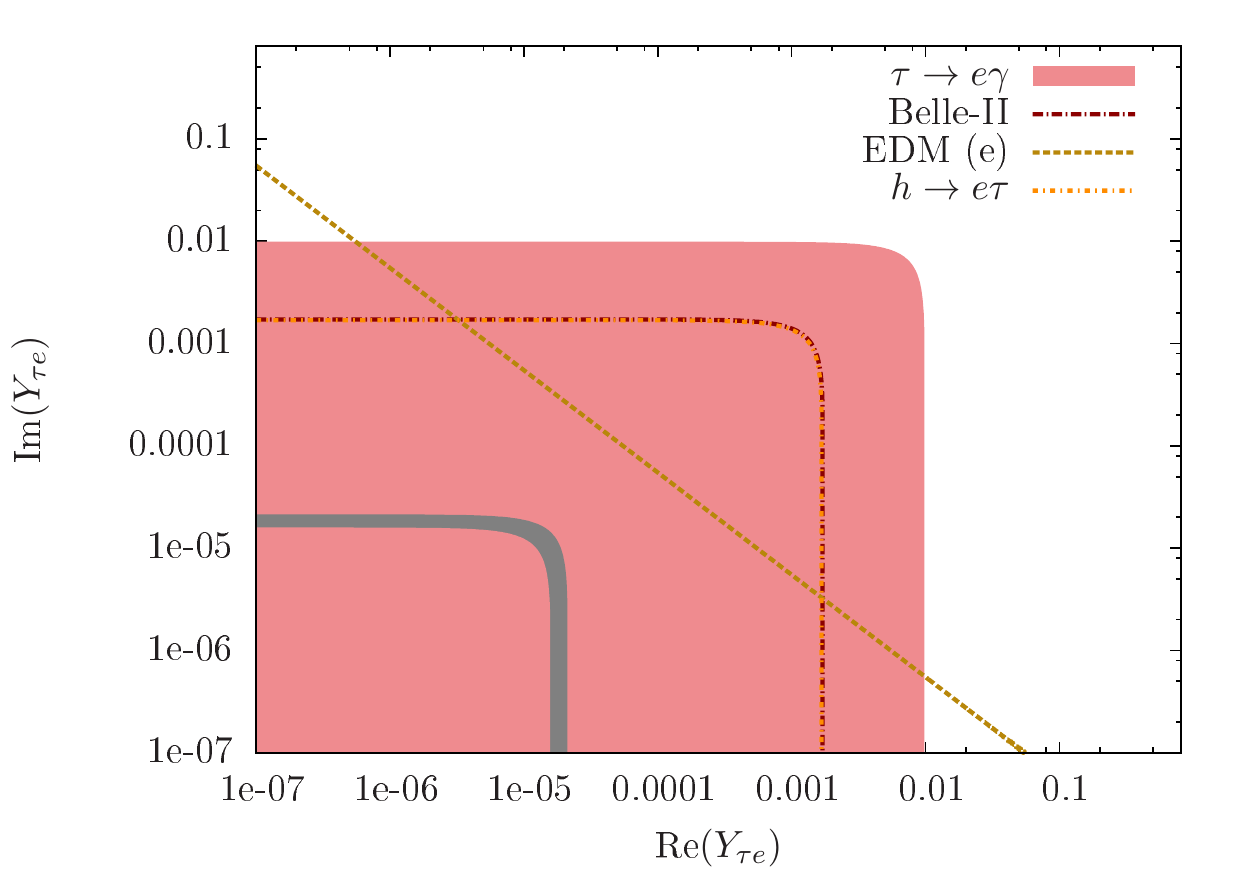}\label{}}
\caption{Figure shows the constraints on the real and imaginary parts of $Y_{\tau e}$. Left panel: The gray region 
(scatter plot) satisfies the flavour violating constraints $\tau \to e \gamma$, $\tau \to 3e$, electron EDM and 
electron $g-2$. The blue (solid) and red (dotted) lines represent the combined constraint from $h \to \tau \mu$ and 
$\mu \to e \gamma$ for BR$(h \to \mu \tau) = 1.51\%, \; 0.84\%$ respectively. Right panel: The pink region is in agreement  with the  experimental  limit from $\tau \to e \gamma$. The red line 
represents the future sensitivity from  Belle-II. The gray region satisfies the combined constraints from $\tau \to e \gamma$ and $h \to \mu \tau$, where the branching ratio of 
$h \to \mu \tau$ varies between $0.84-1.51$\%.}
\label{fig:lfvetau}}
\end{center}
\end{figure}

We summarise the constraints from the direct searches, low energy experiments and sensitivities from the future experiments in Table~\ref{tab:flvsum}. For simplicity, 
we assume that $Y_{\beta\alpha}=Y_{\alpha\beta}$ with $\alpha,\beta = e,\mu, \tau $. We find that the strongest individual constraints on $Y_{\mu \tau}$ and $Y_{e \tau}$ come 
from the run-I searches at the LHC. These constraints however have been obtained assuming no correlation between the two flavour violating couplings. Assuming the validity of 
an EFT and a non-zero $Y_{\mu \tau}$ explaining the excess seen by CMS, the limits on $Y_{e \tau}$ are strongest from a measurement of BR($\mu \to e \gamma$) which sets 
an upper limit on  $|Y_{\mu \tau}Y_{e \tau}|$ at 3.98$\times 10^{-8}$. Thus, combining the excess in $h \to \mu \tau$ with the best-fit branching ratio of 
BR$(h \to \mu \tau)=0.84\%$ and the MEG limit~\cite{Adam:2013mnn} on $\mu \to e \gamma$, we get $Y_{e \tau} \sim 2 \times 10^{-5}$, which is stronger than 
present LHC limit ($Y_{e \tau} < 1.7 \times 10^{-3}$ at 95\% CL). For generic complex Yukawas, we show the constraints on $\textrm{Re}(Y_{e\tau})$ and $\textrm{Im}(Y_{e \tau})$ in Fig.~\ref{fig:lfvetau}.
In the left panel, we show the existing constraints from the direct searches and the low energy experiments. The gray region is in agreement with the 
flavour violating low energy processes, i.e. $\tau \to e \gamma$, $\tau \to 3e$, electron EDM and electron $g-2$ as summarised in Table~\ref{tab:flvsum}. Finally we show the combined constraint from $h \to \mu \tau$ (CMS) and $\mu \to e \gamma$ 
 (MEG). The blue (solid) and red (dashed) lines represent BR($h \to \mu \tau$) = 1.51\% and 0.84\% respectively. {{In this figure, for the charged lepton decays $\tau \to e \gamma$ and $\tau \to 3e$, we have considered the one loop contributions. The two loop contribution for
 $\mu \to e \gamma$ depends on the $Y_{\mu e}$ coupling and vanishes in the limit $Y_{\mu e}=0$. }}
In the right panel, we show the individual limits from $\tau \to e \gamma$ (pink region) and the future sensitivity from 
Belle-II~\cite{Wang:2015kmm} (red line) which will constrain the flavour violating Yukawa couplings by a further factor of $\mathcal{O}(0.1)$. The gray region in the right panel satisfies the 
constraint from $h \to \mu \tau$ and $\mu \to e \gamma$ decays, where the branching ratio of $h \to \mu \tau$ ranges from 
$0.84-1.51\%$. From Fig.~\ref{fig:lfvetau}, it is evident that the limit on $Y_{ \tau e}$ is more stringent than the current
LHC limit, provided that BR$(h \to \mu \tau)$ lies between $0.84-1.51$\% which is required to explain the excess seen by CMS.
In case $Y_{\mu \tau}$ is negligibly small, then BR$(h \to e \tau)$ can be as large as 0.69\% from the direct LHC searches.
The future sensitivity of MEG-II~\cite{Renga:2014xra}  will constrain this coupling even 
further. In addition, the future constraints coming from the Mu$2$e experiment can become even more severe~\cite{Abrams:2012er}.
However, these constraints depend strongly on the validity of the  effective field theory. If the underlying degrees of 
freedoms are light enough, then the EFT description will not be valid \cite{Arganda:2015naa, deLima:2015pqa}.

In the following section ~\ref{sec:3}, we analyse the collider reach to probe lepton-flavour violating Higgs interactions in the $e\mu$, $\mu \tau$ and 
$e\tau$ decay modes for the 14 TeV LHC and its future upgrades with $3000 \, \rm{fb}^{-1}$. Subsequently, we analyse the reach of 
$h\to e \tau$ at the ILC.

\section{Direct detection of LFV at the HL-LHC}
\label{sec:3}
From the previous section, we see that the strongest constraint on $Y_{\mu \tau}$ arises from a direct search at CMS. However,
the constraints from low energy measurements on $Y_{e \tau}$ and $Y_{e \mu}$ are still considerably stronger. However, it is important to note that all constraints derived from the low energy experiments 
are subject to correlations among various Yukawa couplings, which are innate in any EFT approach. In this section, we evaluate the 
high-luminosity LHC's potential to set limits on the three LFV Higgs decays, while being completely agnostic about low-energy 
constraints. If direct searches find evidence for both $h \to \mu \tau$ and $h \to e \tau$ in the near future - 
and if then as a result the bound on $|Y_{\mu \tau} Y_{e \tau}|$ is found to be weaker than obtained from the low energy experiments an interpretation in terms of an effective field theory approach will be at stake and has to be augmented with a less constraining theory assumption.

Taking a cue from Sec.~\ref{sec:leptbound}, here we evaluate the possible reach of a 14 TeV high-luminosity LHC with integrated luminosities up to 3000 fb$^{-1}$ in measuring lepton-flavour violation in the Higgs sector. Hence, we implement 
the relevant flavour violating interactions of the Higgs and the charged leptons in {\tt FeynRules}~\cite{Alloul:2013bka}. The 
generated {\tt Universal FeynRules Output (UFO)}~\cite{Degrande:2011ua} model is then used as input to the Monte-Carlo (MC) event 
generator {\tt MadGraph5 aMC@NLO}~\cite{Alwall:2014hca}. Parton-showering
and hadronisation is carried out using {\tt Pythia 6}~\cite{Sjostrand:2006za}. Thus, the $\tau$ 
decays are simulated using {\tt TAUOLA}~\cite{Jadach:1993hs}. Finally we perform a detector simulations using {\tt Delphes 
3}~\cite{deFavereau:2013fsa} where the jets are reconstructed using the anti-$k_t$ algorithm~\cite{Cacciari:2008gp} with a jet 
parameter of $R=0.5$, as implemented in {\tt FastJet}~\cite{Cacciari:2011ma}. In the following three subsections we show the
prospects of each of the lepton flavour violating Higgs decays.

\subsection{Prospect of $\mu \tau$ channel}
\label{sec:3.1}


In this subsection we focus on the channel where an excess of events have already been seen, \textit{i.e.} in 
$h \to \mu \tau$. Among all the possible final states, e.g. $e \mu + \slashed{E}_T$, $\mu \mu +\slashed{E}_T$  and 
$\mu + \tau_{h} +\slashed{E}_T$, the channel where $\tau \to \mu + \slashed{E}_T$ is the cleanest one. In order to reduce the 
backgrounds, we implement the CMS-like selection cuts as listed in Ref.~\cite{Khachatryan:2015kon}. The  different backgrounds 
for this final state are listed in Table.~\ref{tab:event_mutau_LHC}. The order to which these backgrounds are computed is 
discussed in section~\ref{sec:3.2}. For the signal, we first consider the branching ratio of 
$h \to \mu \tau$ as 0.84$\%$, that corresponds to the central value for the excess reported by CMS. Hence, for the 14 TeV LHC, the signal and background events for $p p \to h \to e \mu +\slashed{E}_T$ with $\mathcal{L} \sim$ 37 
$\rm{fb}^{-1}$ are 251 and 1135 respectively. This results in a 2.1$\sigma$ statistical significance. We further analyse the 
sensitivity reach of the HL-LHC for this channel. In Table~\ref{tab:significance_BR_mutau}, we list the number of signal 
and background events for an integrated luminosity of 3000 fb$^{-1}$. We find that this branching ratio $h \to \mu \tau$ can 
be constrained down to 0.76$\%$ (1.90$\%$) with a 2$\sigma$ (5$\sigma$) statistical significance. However, this can be further 
optimised by adding more cuts or using a multi-variate analysis. We discuss such an optimisation for the $h \to e \tau$ channel
in the section~\ref{sec:3.2}.

To derive the above significances and reach, we define the statistical significance, $\mathcal{S}_1$ by assuming a flat 10\% systematic 
uncertainty on the backgrounds
\begin{equation}
 \mathcal{S}_1 = \frac{S}{\sqrt{S+B+\kappa^2 B^2}},
\end{equation}
where $\kappa=10$\% in our case. We further note that for a very low value of systematic uncertainty ($\kappa \simeq 0$), the significance
is given by
\begin{equation}
 \mathcal{S}_2=S/\sqrt{S+B}
 \label{eq:no-syst}
\end{equation}
and one achieves 2$\sigma$ significance for a branching ratio of $\sim 0.025\%$.
\begin{table}[!ht]
\centering
\begin{tabular}{|c|c|c|c|}
\hline 
Channel   & S(B) (CMS-like) &  $NEV_{sc}^{CMS}$ \\
\hline 
\multirow{4}{2cm}{$e \mu + \slashed{E}_T$}  & Signal & 2421 \\ \cline{2-3} 
&   $\tau \tau+1$ jet                                & 38595 \\
&   $VV$                                             & 18822 \\
&   $W+2$ jets                                       & 6517  \\
&   $t\bar{t}$                                       & 25363 \\
&   single top                                       & 1385 \\
&   SM Higgs                                         & 1319 \\
\hline 
&  \rm{Total background}                             & 92001 \\
\hline
\end{tabular} 
\caption{Signal events for BR$(h \to \mu \tau) = 0.1\%$ after the CMS-like selection
cuts. We also show the corresponding background events for the same set of cuts. The number of events are computed for 
$\mathcal{L}=3000$ fb$^{-1}$.}
\label{tab:event_mutau_LHC}
\end{table}

\begin{table}[!ht]
\centering
\begin{tabular}{|c|c|}
\hline 
Channel   & BR \% ($\mathcal{S}^{CMS}$) \\
\hline 
\multirow{2}{2cm}{$e \mu + \slashed{E}_T$}  & 0.76 ($2\sigma$) \\ \cline{2-2} 
&   1.90 ($5\sigma$) \\
\hline
\end{tabular} 
\caption{The lowest branching ratios BR$(h \to \mu \tau)$ that can be probed at $2\sigma$ and $5\sigma$ significance at the 
14 TeV LHC with $\mathcal{L}=3000$ fb$^{-1}$.}
\label{tab:significance_BR_mutau}
\end{table}

\subsection{Prospect of $e\tau$ channel}
\label{sec:3.2}

Here we consider the flavour violating Higgs decay to $e\tau$, followed by the hadronic as well as leptonic decays 
of $\tau$. For the 14 TeV $e\tau$ analysis we follow the proposal of Ref.~\cite{CMS:2011msa} and adopt a $\tau$ tagging and 
mistagging efficiencies of (40\%, 0.2\%)\footnote{Before performing the 14 TeV analysis for these three LFV decays, we validated the 8 TeV $p p \to h \to \mu \tau_{had}$ numbers as reported in the CMS run-I~\cite{Khachatryan:2015kon} results.}. Here we consider the following final 
states: 

\begin{itemize}
\item
$ p p \to h \to e \tau \to e e + \slashed{E}_T$
\item
$ p p \to h \to e \tau \to \mu e + \slashed{E}_T$
\item
$ p p \to h \to e \tau \to e \tau_{had} + \slashed{E}_T$
\end{itemize}

The major SM backgrounds for the processes mentioned above are $\tau \tau+$ jet (computed at the next-to-next-to leading order 
(NNLO)~\cite{Boughezal:2016isb}), 
$VV$ (with $V = W^{\pm},Z$) (at the next-to leading order (NLO)~\cite{Campbell:2011bn}), $W+$ jets (at NLO~\cite{Campbell:2003hd}), 
with $W$ decaying leptonically and one of the jets mistagged as a $\tau$-hadron, $ee+$jets (computed at NNLO~\cite{Catani:2009sm}), 
$t\bar{t}$ (at next-to-next-to-next-to leading order (N$^3$LO)~\cite{Muselli:2015kba}), single-top (at NLO~\cite{Kant:2014oha}) and
the SM-Higgs backgrounds (also computed at NNLO), i.e. $h \to \tau^+ \tau^-$. For the single-top background, the $Wt$ mode has the dominant 
contribution for our scenario, whereas the $s$- and $t$-channel contributions are negligible. So, for simplicity, we multiplied our leading order (LO) cross-section
by the NLO $k$-factor for the $Wt$ mode. Besides these, there are some fake backgrounds
like QCD multi-jets, where the jets might fake leptons, and $W\gamma$, where the photon might convert to an electron-positron
pair. However, it is very difficult to get a proper estimate for these backgrounds without doing a full detector simulation
or without using data driven methods. For all practical purposes, these backgrounds will not significantly alter our quoted results.
Hence we neglect these in the present analysis.

For the $\tau \tau+$ jet background, \textit{viz.}, the major background for the $e\mu$ final state, we perform an 
ME-PS MLM matching in the MadGraph framework. For the $W+$ jets background, which is the single most important background for 
the $e\tau_{had}$ final state, we also simulate a matched sample with up to two partonic jets. We select only those events where at least 
one of the jets fake a $\tau$-hadron. We perform an inclusive study and demand no jet veto. Similarly a matched sample was 
generated for the $ee+$jet background, which is the dominant background for the $ee$ final state. 
To optimise the signal sensitivity, we adopt similar cuts as done by the CMS 8 TeV analysis for $h \to e \tau$ final 
state~\cite{CMS:2015udp}. In addition, we also optimise over the $p_T$ cuts in order to gain maximum sensitivity.

The common set of trigger cuts that we use for all the final states in this subsection are:
\begin{itemize}
\item
The transverse momentum of $e, \mu$ and jet are: $p_T(e) > 10 \, \rm{GeV}$, $p_T(\mu) > 10 \, \rm{GeV} $ and 
$p_T(j) > 30 $ GeV respectively,
\item
The pseudo-rapidity of $e, \mu$ and jet: $|\eta(e)| < 2.1$, $|\eta(e)| < 2.3$ and $|\eta(j)| < 4.7$.
\end{itemize}

We use different selection cuts for the three different final states $ee + \slashed{E}_T$, $\mu e + \slashed{E}_T$ and 
$e \tau_{had} + \slashed{E}_T$. For the $e\mu+\slashed{E}_T$ final state we use the following selection cuts:
\begin{itemize}
 \item $\slashed{E}_T > 30$ GeV
 \item The azimuthal angle separations: $\Delta \phi_{\vec{\mu}-\vec{\slashed{E}}_T} < 0.5$ and $\Delta \phi_{\vec{\mu}-\vec{e}} > 2.7$
 \item The transverse mass variable: $M_T(\mu)<65$ GeV and $M_T(e)>50$ GeV, where the transverse mass is defined as
       \begin{equation}
          M_T(\ell) = \sqrt{2 p_T(\ell)\vec{\slashed{E}}_T (1 - \cos \Delta \phi_{\vec{\ell}-\vec{\slashed{E}}_T})}
       \end{equation}
 \item The collinear mass variable: $105$ GeV $< M_{collinear}^{\mu e} <  145$ GeV, where the collinear mass is the following,
       \begin{equation}
        M_h = M_{collinear} = \frac{M_{vis}}{\sqrt{x_{\tau_{vis}}}},
       \end{equation}
       with the visible momentum fraction of the $\tau$ decay products being,
       $x_{\tau_{vis}}=\frac{|\vec{p}_T^{\; \tau_{vis}}|}{|\vec{p}_T^{\; \tau_{vis}}|+|\vec{p}_T^{\; \nu}|}$, where
       $\vec{p}_T^{\; \nu}=|\vec{\slashed{E}}_T| \hat{p}_T^{\; \tau_{vis}}$
 \item In addition,  we use 10 sets of cuts for $p_T^e$ and $p_T^{\mu}$ optimised around the CMS-like cut, \textit{viz.} 
       $p_T^e > 50$ GeV and $p_T^{\mu} > 10$ GeV.
\end{itemize}

For the $ee+\slashed{E}_T$ final state, the cuts are exactly same as in the previous case with the following transformations $
\mu \to e_2, e \to e_1$, where $e_1$ is the electron coming from the Higgs decay and $e_2$ comes from the $\tau$ decay. 
For the remaining final state $e\tau_{had} + \slashed{E}_T$, we use the following selection cuts:
\begin{itemize}
 \item $\slashed{E}_T < 40$ GeV
 \item Azimuthal angle separation: $\Delta \phi_{\vec{e}-\vec{\tau}_{had}} > 2.7$ and transverse mass $M_T(\tau_{had})<50$ GeV
 \item The collinear mass: $105$ GeV $< M_{collinear}^{e \tau_{had}} <  145$ GeV
 \item In addition, here also we use 10 sets of cuts for $p_T^e$ and $p_T^{\tau_{had}}$ optimised around the CMS-like cut 
    $p_T^e > 40$ GeV and $p_T^{\tau_{had}} > 35$ GeV.
\end{itemize}

In addition to the above CMS-like selection cuts, we implement the cut on $\slashed{E}_T$ and the optimised $p_T$ cut giving 
the maximum sensitivity (shown in Table~\ref{tab:ptopt}). We show the number of events after all the trigger and selection 
cuts in Table~\ref{tab:event_CB} and show the exclusion limit of the BR$(h \to e\tau)$ in Table~\ref{tab:significance_BR}.  

\begin{table}[!ht]
\centering
\begin{tabular}{|c|c|}
\hline 
Channel   & optimal $p_T$ cut \\
\hline
$e \mu+ \slashed{E}_T$  &  $p_T^e > 50$ GeV and $p_T^{\mu} > 10$ GeV \\
$ee+ \slashed{E}_T$  & $p_T^{e_1} > 50$ GeV and $p_T^{e_2} > 10$ GeV \\
$e \tau_{had}+ \slashed{E}_T$  & $p_T^e > 55$ GeV and $p_T^{\tau_{had}} > 50$ GeV \\
\hline
\end{tabular} 
\caption{Optimized $p_T$ cuts for the three final states.}
\label{tab:ptopt}
\end{table}

\begin{table}[!ht]
\centering
\begin{tabular}{|c|c|c|c|}
\hline 
Channel   & S(B) (optimal) &  $NEV_{sc}^{optimal}$ \\
\hline 
\multirow{4}{2cm}{$e \mu + \slashed{E}_T$}  & Signal & 1600 \\ \cline{2-3} 
&   $\tau \tau+1$ jet & 21161 \\
&   $VV$              & 7179 \\
&   $W+2$ jets        & 6517 \\
&   $t\bar{t}$        & 12455 \\
&   single top        & 923 \\
&   SM Higgs          & 466 \\
\hline 
&  \rm{Total background}                             & 48701 \\
\hline
\hline
\multirow{4}{2cm}{$e e + \slashed{E}_T$}  & Signal & 1291 \\ \cline{2-3} 
&   $\tau \tau+1$ jet & 16636 \\
&   $VV$              & 19135 \\
&   $ee+1$ jet        & 17061 \\
&   $t\bar{t}$        & 8605 \\
&   single top        & 1077 \\
&   SM Higgs          & 310 \\
\hline
&  \rm{Total background}                             & 62824 \\
\hline
\hline
\multirow{4}{2cm}{$e \tau_{had} + \slashed{E}_T$}  & Signal & 1013 \\ \cline{2-3} 
&   $\tau \tau+1$ jet & 11578 \\
&   $VV$              & 2372 \\
&   $W+2$ jets        & 81465 \\
&   $ee+1$ jet        & 4981 \\
&   $t\bar{t}$        & 2038 \\
&   single top        & 1693 \\
&   SM Higgs          & 388 \\   
\hline 
&  \rm{Total background}                             & 104515 \\
\hline
\end{tabular} 
\caption{Signal events for BR$(h \to e\tau) = 0.1\%$ after all selection
cuts. The superscript ``$optimal$'' signifies the number of events for the optimal $p_T$ cuts. We also show the corresponding 
background events for the same set of cuts. The number of events  are computed for $\mathcal{L}=3000$ fb$^{-1}$.}
\label{tab:event_CB}
\end{table}

\begin{table}[!ht]
\centering
\begin{tabular}{|c|c|}
\hline 
Channel   & BR \% ($\mathcal{S}^{optimal}$) \\
\hline 
\multirow{2}{2cm}{$e \mu + \slashed{E}_T$}  & 0.61 ($2\sigma$) \\ \cline{2-2} 
&   1.53 ($5\sigma$) \\
\hline 
\multirow{2}{2cm}{$e e + \slashed{E}_T$}  & 0.97 ($2\sigma$) \\ \cline{2-2} 
&   2.44 ($5\sigma$) \\
\hline 
\multirow{2}{2cm}{$e \tau_{had} + \slashed{E}_T$}  & 2.06 ($2\sigma$) \\ \cline{2-2} 
&   5.17 ($5\sigma$) \\
\hline
\end{tabular} 
\caption{The lowest branching ratios BR$(h \to e \tau)$ that can be probed at $2\sigma$ and $5\sigma$ C.L.}
\label{tab:significance_BR}
\end{table}
From Table~\ref{tab:significance_BR}, one can observe a BR$(h \to e \tau)$ of around 1.5\% at $5\sigma$
for the $e\mu$ final state. For the $ee$ final state we require a somewhat larger branching ratio to have a $5\sigma$ 
statistical significance. Whereas for the hadronic final state one can not go below $\sim 5.1$\% in order to obtain
a $5\sigma$ significance with such a cut-based analysis. If we statistically combine these three significances in quadrature, 
then we attain a $2\sigma$ significance for a branching ratio of 0.50\%. Combining them additively, one achieves $2\sigma$ 
for as low a branching ratio as $\sim 0.32$\%. To see if one can probe lower branching ratios in the $e\mu$
channel, we exploit the kinematics of both the signal and the background in more details. We perform a multivariate analysis
with the Boosted Decision Tree (BDT) algorithm using the root based TMVA \cite{TMVA2007} framework. We choose 11 kinematic 
variables for this purpose, \textit{viz.} 
\begin{align}
|\vec{p}_T^{\; e}|, \; |\vec{p}_T^{\; \tau_{had}}|, \; \Delta \phi_{\vec{e}-\vec{\slashed{E}_T}}, \; 
\Delta \phi_{\vec{\tau_{had}}-\vec{\slashed{E}_T}}, \; \Delta \phi_{\vec{e}-\vec{\tau_{had}}},\nonumber \\
\; M_T(e), \; M_T(\tau_{had}),\; M_{e\tau_{had}}^{vis.}, \; M_{collinear}^{e \tau_{had}}, \; \slashed{E}_T, \; \phi_{\vec{\slashed{E}}_T}. \nonumber
\end{align}

For the MVA, we take care of the issue of overtraining of the signal/background. The Kolmogorov Smirnov
(KS) test is used to check for the overtraining of a sample. The test sample
is not overtrained if the KS probability lies between 0.1 to 0.9. In most cases, a critical KS
probability value more than 0.01 \cite{KS-test} ensures that the signal and background samples are not overtrained.

We find after a proper training of the sample, that an optimised cut on the BDT yields a better reach on the branching ratio.
In table \ref{tab:event_BDT}, we tabulate the signal and dominant background events at an integrated luminosity of 3000 fb$^{-1}$ after
an optimised cut on the BDT variable.

\begin{table}[!ht]
\centering
\begin{tabular}{|c|c|c|c|}
\hline 
Channel   & S(B) (optimal) &  $NEV_{BDT}^{optimal}$ \\
\hline 
\multirow{4}{2cm}{$e \mu + \slashed{E}_T$}  & Signal & 277 \\ \cline{2-3} 
&   $\tau \tau+1$ jet                                & 3859 \\
&   $VV$                                             & 936 \\
&   $t\bar{t}$                                       & 1585 \\
&   single top                                       & 197 \\
\hline 
&  \rm{Total background}                             & 6577 \\
\hline
\end{tabular} 
\caption{Same as in Table~\ref{tab:event_CB} for the $h \to e \mu + \slashed{E}_T$ channel after an optimal cut on the
BDT variable.}
\label{tab:event_BDT}
\end{table}

We find that one can go down to as low as $\sim 0.48\%$ in order to achieve a 2$\sigma$ significance. To achieve a 5$\sigma$ 
discovery one can not go below a branching ratio of 1.20\%. We see that the MVA analysis improves the reach by a factor of 
$\sim 1.28$ for the $e \mu + \slashed{E}_T$ final state. We also note that in  \ref{sec:3.2} we obtained the reach on BR($h \to \mu \tau$) 
with the CMS-like cuts. With an MVA, however, we expect a similar improvement in this channel as in the $e \tau$ sector.

We also note that one can attain a 2$\sigma$ significance in the $e \mu + \slashed{E}_T$ channel with the 
cut-based analysis for a branching ratio of 0.028\% by using Eq.~\ref{eq:no-syst}.

\subsection{Prospect of $e \mu$}
\label{sec:3.3}
Inspired by CMS's direct search for the flavour violating decay $h \to e \mu$~\cite{CMS:2015udp}, we study the prospect of observing this channel
at the HL-LHC. For this analysis, we apply the following simple set of cuts :

\begin{itemize}
 \item $p_T(e) > 40$ GeV and $p_T(\mu) > 40$ GeV
 \item $|\eta_e| < 1.479$ and $|\eta_{\mu}| < 0.8$ (in the barrel)
 \item $\slashed{E}_T < 20$ GeV
 \item $123 \textrm{ GeV } < m_h < 127 \textrm{ GeV}$.
\end{itemize}

Here also we compute the backgrounds at the orders specified in section~\ref{sec:3.2}. The major backgrounds are 
$e \mu + \slashed{E}_T$ (dominantly from $WW$ production), Drell-Yan production of $\tau \tau$, $t\bar{t}$,
$e(\mu) + \tau + \slashed{E}_T$ (also dominantly from $WW$ production), $\tau \tau + \slashed{E}_T$ (mainly coming from $WW$ and $ZZ$) and single top (dominantly
from the $Wt$ production). In Table~\ref{tab:event_emu_LHC} we list the number of signal and background events after all the
selection cuts for the HL-LHC at 3000 fb$^{-1}$. Finally in Table~\ref{tab:significance_BR_emu}, we show the prospective reach.

\begin{table}[!ht]
\centering
\begin{tabular}{|c|c|c|c|}
\hline 
Channel   & S(B) &  $NEV_{sc}^{optimal}$ \\
\hline 
\multirow{4}{0.5cm}{$e \mu$}  & Signal                         & 1435 \\ \cline{2-3} 
&   $e \mu + \slashed{E}_T$                                    & 2449 \\
&   $\tau \tau$                                                & 406 \\
&   $t\bar{t}$                                                 & 9511 \\
&   $e(\mu) + \tau + \slashed{E}_T$                            & 152 \\
&   $\tau \tau + \slashed{E}_T$                                & 5 \\
&   single top                                                 & 1231 \\
\hline 
&  \rm{Total background}                                       & 13754 \\
\hline
\end{tabular} 
\caption{Signal events for BR$(h \to e \mu) = 0.01\%$ after the optimised selection
cuts. We also show the corresponding background events for the same set of cuts. The number of events are computed for 
$\mathcal{L}=3000$ fb$^{-1}$.}
\label{tab:event_emu_LHC}
\end{table}

\begin{table}[!ht]
\centering
\begin{tabular}{|c|c|}
\hline 
Channel   & BR \% ($\mathcal{S}$) \\
\hline 
\multirow{2}{0.5cm}{$e \mu$}  & 0.0193 ($2\sigma$) \\ \cline{2-2} 
& 0.0482 ($5\sigma$) \\
\hline
\end{tabular} 
\caption{The lowest branching ratios BR$(h \to e \mu)$ that can be probed at $2\sigma$ and $5\sigma$ significance at the 
14 TeV LHC with $\mathcal{L}=3000$ fb$^{-1}$.}
\label{tab:significance_BR_emu}
\end{table}

Here also we note that one can attain a 2$\sigma$ significance in the $e \mu$ channel with the 
this analysis for as low a branching ratio of $1.65\times 10^{-3}\%$ by using Eq.~\ref{eq:no-syst}.
\section{$h \to e \tau$ at ILC}
\label{sec:4}

In the previous section, we discussed the prospects of observing a lepton flavour violating Higgs in all the three possible
channels at a 14 TeV LHC with an integrated luminosity going up to 3000 fb$^{-1}$. We saw that one can definitely expect
improvements compared to the 8 TeV results but due to the large backgrounds and huge uncertainties, these are not so dramatic 
as one would like.
We know that LHC is plagued with huge backgrounds and hence we can expect better precision at lepton colliders. In this 
section, we repeat the analysis for $h \to e \tau$ for centre of mass energies of $\sqrt{s}=250$ GeV and 1 TeV at an ILC 
machine. Here we just want to point out the improvement over the 
LHC. A similar improvement can be expected for the $h \to \mu \tau$ as well. For $e \mu$, we do not expect
a significant improvement at ILC because in section~\ref{sec:3.3}, we already saw that the reach for the branching ratio can
be as low as $1.95 \times 10^{-2}$\%. The number of signal events are expected to be very low at the ILC for such small
branching ratios.

The two main topologies that we study here are the associated production of the Higgs with a $Z$-boson and the Higgs produced 
in association with neutrinos through the $t$-channel fusion diagram. At 250 GeV, the associated production with a $Z$-boson 
offers the largest cross-section with the $Z$ decaying hadronically, whereas for the leptonic modes of $Z$, 
the total cross-section is suppressed. Hence, we study in detail the $Z h, Z \to q \bar{q}$ production for $\sqrt{s}=250$ GeV. 
For the 1 TeV study, we consider the $\slashed{E} h$ channel which includes both the $t$-channel contribution mediated by 
$W$-boson and an $s$-channel contribution from the $Zh$ topology. However, here the contribution coming from the latter is 
nominal. The different final states that we study in this section can be summarised as :

\begin{itemize}
 \item $e^+ e^- \to Z h, h \to \tau e$, with $Z  \to \ge 2j$ and 
       $\tau \to e \nu, \mu \nu$ or $\tau$ tagged as $\tau_{had}$
 \item $e^+ e^- \to \slashed{E} h, h \to \tau e$, with $\tau \to e \nu, \mu \nu$ or $\tau$ tagged as $\tau_{had}$.
\end{itemize}

\subsection{ILC at $\sqrt{s} = 250$ GeV}
\label{sec:4.1}

In this subsection, we simulate a prospective analysis at the ILC with $\sqrt{s}=250$ GeV. The dominant backgrounds for the various final states are: 

\begin{itemize}
\item
$Z \ell \ell \nu \nu$,   $Z \ell \ell \ell \ell $ and $W \ell \nu \ell \ell$, followed by the hadronic decays of $Z $ and $W$, \textit{viz.} 
$Z \to q \bar{q}$ and $W \to q q^{\prime}$. In short, we denote these backgrounds with the tag, $3V$-like,
\item
$e e Z$ and $\tau \tau Z$, followed by $Z \to q q$. In short, we denote these backgrounds as $2V$-like,
where $q =$ light jets $+ \, b$-jets and $\ell = e, \mu, \tau$.
\end{itemize}

Here we perform a cut-based analysis with the following set of general selection cuts:

\begin{itemize}
 \item Demand that at least two jets reconstruct the $Z$-boson mass with a window 
       of 25 GeV, i.e., $M_{Z}-25 \; \textrm{GeV} < M_{jj} < M_Z+25 \; \textrm{GeV}$.
\item
For the associated production $Z h$, we apply the cuts on the invariant mass of the visible products:
 \begin{itemize}
  \item The visible invariant mass lies in 110 GeV $< M_{e\mu} <$ 140 GeV for $e \mu$.
  \item The visible invariant mass lies in 120 GeV $< M_{ee} <$ 130 GeV for $e e$.
  \item The visible invariant mass lies in 110 GeV $< M_{e\tau_{had}} <$ 140 GeV for $e \tau_{had}$.
 \end{itemize}
\item In addition to the above cuts, we demand a cut on $|\cos(\theta_{jj})| < 0.8$ for the $e e j j$ channel in order to reduce
      the $2V$-like background.
\end{itemize}

In Table~\ref{tab:ILC-NEV}, we list the number of signal and background events after the selection cuts for the dominant modes. 
Finally in Table~\ref{tab:ILC-250-sig}, we show the reach of the ILC for BR$(h \to e \tau)$ in these dominant final states with 
$\sqrt{s}=250$ GeV. For this purpose, we use a different formula for the significance as compared to LHC, owing to the fact 
that the systematic uncertainties are expected to be significantly less for the ILC. We use the standard formula as quoted in 
the ILC Snowmass report~\cite{Asner:2013psa}, \textit{viz.}

\begin{equation}
 \mathcal{S} = \frac{S}{\sqrt{S+B}}
\end{equation}

\begin{table}[!ht]
\centering
\begin{tabular}{|c|c|c|c|}
\hline 
Channel   & S(B) (optimal) &  $NEV_{sc}^{optimal}$ \\
\hline 
\multirow{2}{4cm}{$e + \mu + \ge 2 j + \slashed{E}$}  & Signal & 11 \\ \cline{2-3} 
& $3V-$like  & 14 \\
& $2V-$like & 1 \\
\hline 
&  \rm{Total background}                             & 15 \\
\hline
\hline
\multirow{2}{4cm}{$2 e + \ge 2 j + \slashed{E}$}  & Signal & 8 \\ \cline{2-3} 
& $3V-$like  & 4 \\
& $2V-$like & 190 \\
\hline 
&  \rm{Total background}                             & 194 \\
\hline
\hline
\multirow{2}{4cm}{$e +  \tau_{had} + \ge 2 j + \slashed{E}$}  & Signal & 24 \\ \cline{2-3} 
& $3V-$like  & 1 \\
& $2V-$like & 17 \\
\hline 
&  \rm{Total background}                             & 18 \\
\hline
\end{tabular} 
\caption{Signal events for BR$(h \to e\tau) = 1\%$ and the background events after the optimised selection
cuts. The above quantities are computed for $\sqrt{s}=250$ GeV and $\mathcal{L}=250$ fb$^{-1}$.}
\label{tab:ILC-NEV}
\end{table}

\begin{table}[!ht]
\centering
\begin{tabular}{|c|c|}
\hline 
Channel   & BR \% ($\mathcal{S}^{optimal}$) \\
\hline 
\multirow{2}{4cm}{$e + \mu + \ge 2 j + \slashed{E}_T$}  & 0.96 ($2\sigma$) \\ \cline{2-2} 
&  3.39 ($5\sigma$) \\
\hline
\multirow{2}{4cm}{$2 e + \ge 2 j + \slashed{E}_T$}  & 3.93 ($2\sigma$) \\ \cline{2-2} 
&   $>10$ ($5\sigma$) \\
\hline
\multirow{2}{4cm}{$e + \tau_{had} + \ge 2 j + \slashed{E}_T$}  & 0.44 ($2\sigma$) \\ \cline{2-2} 
&   1.54 ($5\sigma$) \\
\hline
\end{tabular} 
\caption{Reach of BR$(h \to e \tau)$ at $2\sigma$ and $5\sigma$ at the ILC with $\sqrt{s}=250$ GeV and $\mathcal{L}=250$ fb$^{-1}$.}
\label{tab:ILC-250-sig}
\end{table}

On statistically combining these three significances in quadrature, we attain a $2\sigma$ significance for a branching ratio 
of 0.38\%. Combining them additively, one achieves $2\sigma$ for as low a branching ratio as $\sim 0.25$\%. 
So we see that the ILC at $\sqrt{s}=250$ GeV performs comparably with the high luminosity LHC for these sets of simple 
cuts. To see if we have better prospects at the 1 TeV ILC, we study the vector boson fusion topology in details in 
section~\ref{sec:4.2}.

\subsection{ILC at $\sqrt{s} = 1$ TeV}
\label{sec:4.2}

The ILC at 1 TeV motivates us to study the Higgs in association with missing energy because of much cleaner backgrounds. Another
reason for studying the prospects at the 1 TeV ILC is the 1000 fb$^{-1}$ integrated luminosity. Here, we implement the 
following selection cuts:  

\begin{itemize}
\item
For the $e \mu + \slashed{E}$ channel, we apply :
\begin{itemize}
 \item $1.5 < \Delta \phi_{\mu \slashed{E}} < 3.0$
 \item The visible momentum, $|\vec{p}_{vis}| < 200$ GeV
 \item $2.4 < \Delta R_{e \slashed{E}} < 4.0$
 \item $|\cos(\theta_{e \mu}) < 0.8|$
 \item The invariant mass of the visible particles, $50 \, \textrm{GeV} \, < M_{e \mu} < 120 \, \textrm{GeV}$
\end{itemize}
\item
For the $ee + \slashed{E}$ channel, we apply :
\begin{itemize}
 \item $2.0 < \Delta \phi_{e_1 \slashed{E}} < 3.0$, where $e_1$ is the electron with the hardest $p_T$
 \item $|\vec{p}_{vis}| < 200$ GeV
 \item $2.0 < \Delta R_{e_1 \slashed{E}} < 4.5$
 \item $\Delta R_{e_1 e_2} < 2.8$, where $e_2$ is the second hardest electron.
 \item $|\cos(\theta_{e_1 e_2}) < 0.8|$
 \item $100 \, \textrm{GeV} \, < M_{e_1 e_2} < 120 \, \textrm{GeV}$
\end{itemize}
For the $e \tau_{had} + \slashed{E}$ channel, we apply :
\begin{itemize}
 \item $1.5 < \Delta \phi_{\tau_{had} \slashed{E}} < 3.0$, where $e_1$ is the electron with the hardest $p_T$
 \item $|\vec{p}_{vis}| < 200$ GeV
 \item $2.0 < \Delta R_{\tau_{had} \slashed{E}} < 4.5$
 \item $60 \, \textrm{GeV} \, < M_{e \tau_{had}} < 130 \, \textrm{GeV}$,
\end{itemize}
where the notations of the variables are self-explanatory.
\end{itemize}

The dominant backgrounds for these channels can be summarised as:
\begin{itemize}
 \item $\tau^{+} \tau^{-}$,
 \item $2 \ell + 2 \nu$ and 
 \item $4 \ell$,
\end{itemize}
where $\ell = e, \mu, \tau$.

These optimised cuts are applied to compute the significance in these three channels. We must note here that the $\Delta \phi$
distributions for the $\tau^+ \tau^-$ background peak at 0 and $\pi$. Hence our $\Delta \phi$ cuts almost completely eradicate
this background. The signal and background events after imposing the above sets of selection cuts are found in 
Table~\ref{tab:ILC-NEV-1TeV}. The significances and the reach are summarised in table~\ref{tab:ILC-1000-sig}.

\begin{table}[!ht]
\centering
\begin{tabular}{|c|c|c|c|}
\hline 
Channel   & S(B) (optimal) &  $NEV_{sc}^{optimal}$ \\
\hline 
\multirow{2}{4cm}{$\mu + e + \slashed{E}$}  & Signal &  86 \\ \cline{2-3} 
& $\tau^{+} \tau^{-}$  &  0 \\
& $2 \ell + 2 \nu$ & 411 \\
& $4\ell + 2 \nu$ & 286 \\
\hline 
&  \rm{Total background}                             & 697 \\
\hline
\hline
\multirow{2}{4cm}{$2 e + \slashed{E}$}  & Signal & 22 \\ \cline{2-3} 
& $\tau^{+} \tau^{-}$  &  0 \\
& $2 \ell + 2 \nu$ & 521 \\
& $4\ell + 2 \nu$ & 336 \\
\hline 
&  \rm{Total background}                             & 857 \\
\hline
\hline
\multirow{2}{4cm}{$e + \tau_{had} + \slashed{E}$}  & Signal & 312 \\ \cline{2-3} 
& $\tau^{+} \tau^{-}$  &  1 \\
& $2 \ell + 2 \nu$ & 1028 \\
& $4\ell + 2 \nu$ & 243 \\
\hline 
&  \rm{Total background}                             & 1272 \\
\hline
\end{tabular} 
\caption{Signal events for BR$(h \to e\tau) = 1\%$ and the background events after the optimised selection
cuts. The above quantities are computed for $\sqrt{s}=1$ TeV and $\mathcal{L}=1000$ fb$^{-1}$.}
\label{tab:ILC-NEV-1TeV}
\end{table}

\begin{table}[!ht]
\centering
\begin{tabular}{|c|c|}
\hline 
Channel   & BR \% ($\mathcal{S}^{optimal}$) \\
\hline 
\multirow{2}{4cm}{$e + \mu + \slashed{E}$}  & 0.63 ($2\sigma$) \\ \cline{2-2} 
&  1.68 ($5\sigma$) \\
\hline
\multirow{2}{4cm}{$2 e + \slashed{E}$}  & 2.75 ($2\sigma$) \\ \cline{2-2} 
&  7.22 ($5\sigma$) \\
\hline
\multirow{2}{4cm}{$e + \tau_{had} + \slashed{E}$}  & 0.24 ($2\sigma$) \\ \cline{2-2} 
&  0.62 ($5\sigma$) \\
\hline
\end{tabular} 
\caption{Reach of BR$(h \to e \tau)$ at $2\sigma$ and $5\sigma$ at the ILC with $\sqrt{s}=1$ TeV and 
$\mathcal{L}=1000$ fb$^{-1}$.}
\label{tab:ILC-1000-sig}
\end{table}

Here also, on statistically combining the three significances in quadrature, 
we obtain a $2\sigma$ significance for a branching ratio of 0.22\%. Combining them additively, one achieves $2\sigma$ 
for as low branching a ratio as $\sim 0.16$\%. We see that the ILC at 1 TeV with $\mathcal{L}=1000 \, \textrm{fb}^{-1}$
performs better. But, we must note that with neither scenario we can measure a branching ratio of less than 0.1\%.

In doing the computations for the ILC, we consider unpolarised beams for the 250 GeV analysis because there is no
significant enhancement in the signal. But for the 1 TeV analysis we consider the $e^-$ and $e^+$ polarisations 
as (-80,20). Here we see a significant increase in the signal cross-section by $\sim 2.15$ times even though the 
dominant background, i.e. $2 \ell 2 \nu$ also increases by a factor of 2. So we gain in sensitivity for
the polarised beams in this scenario. In contrast to the LHC studies, here we have performed a leading order
analysis. However, the next-to-leading order effects are not expected to change our conclusions appreciably.

Here, we must comment on the fact that the ILC capabilities~\cite{Peskin:2013xra} on measuring an invisible branching ratio (which can very well
be read as non-standard branching ratio) can be as low as 0.54\% for a 250 GeV ILC machine with an integrated luminosity of
250 fb$^{-1}$ and this decreases to around 0.22\% for $\sqrt{s} = 1$ TeV with $\mathcal{L} = 1000$ fb$^{-1}$. As we can see
that the numbers that we have obtained are in the ballpark of these quoted limits. We also mention that our 
analysis in the ILC sector can be further improved by optimising the cuts to a greater degree or by using a multivariate
technique after identifying interesting variables.
\section{Summary}
\label{sec:5}

The run-I results from CMS and ATLAS gave us the first hint at flavour violation in the Higgs sector in the channel 
$h \to \mu \tau$ with a best-fit branching ratio of less than a percent. The 8 TeV collider searches in the other two LFV 
channels, i.e. $h \to e \tau$ and $h \to e \mu$  did not have any significant excess over the respective SM 
backgrounds. Confirmation of any of these flavour violating processes will necessarily indicate the existence of new physics. 

In this work, we performed rigorous collider analyses to get an estimates of the discovery prospect of all the three lepton 
flavour violating Higgs decays, at the high luminosity run of the 14 TeV LHC.  We also performed a case study for 
$h \to e \tau$ at the ILC with $\sqrt{s}=250$ GeV and 1 TeV. 

Below we summarise our findings from the collider analyses.

\begin{itemize}

\item
We analysed the prospect of $h \to \mu \tau$ and $h \to e \tau$ at the 14 TeV LHC with $\mathcal{L} = 3000$ fb$^{-1}$ and found that 
a BR$(h \to \mu \tau / e \tau )$ of $\sim$ 0.5 $\%$ can be probed with a 2$\sigma$ significance.   

\item We obtained the prospects of observing the cleaner channel $h \to e \mu$ at the HL-LHC with an
integrated luminosity of 3000 fb$^{-1}$. We found that to achieve a 2$\sigma$ significance, one can go as low as $1.95\times 10^{-2}\%$ in the branching ratio.

\item
In addition to the LHC, we also showed the prospects of studying Higgs flavour violation at the ILC with $\sqrt{s}=250$ GeV and 
1 TeV. A branching ratio, BR$(h \to e \tau)$ as low as 0.24\% can be probed for the the $h + \slashed{E}$ channel with a 2$\sigma$ statistical significance 
and with an integrated luminosity of 1000 fb$^{-1}$ at the 1 TeV ILC. We also expect similar improvement in the $\mu \tau$ sector compared to HL-LHC. 
\end{itemize}

Hence, we see that the discovery of lepton flavour violation in $h \to \mu \tau$ and/or the $h \to e \tau$ at collider experiments 
will imply large branching ratios of few times $\mathcal{O}(0.1\%)$. The simultaneous discovery of both these channels 
at the 14 TeV HL-LHC or at the ILC will no doubt question the validity of an EFT approach. This will prompt
us in building models with relatively light massive states. However, if there is only evidence for one of the processes, we can not make strong statements about the existence of light degrees of freedom. Our present study thus encourages the experimental groups to look for these unique signatures and gives an estimate of the smallest branching ratios that can be probed at the near-future colliders.

\acknowledgments
We thank C\'{e}dric Delaunay, Shilpi Jain, Tanumoy Mandal and Emanuele Re for useful discussions and technical help during various phases
of this work. SB acknowledges the support of the Indo French LIA THEP (Theoretical high Energy Physics) of the CNRS. The work of BB is supported by Department of Science and 
Technology, Government of INDIA under the Grant Agreement numbers IFA13-PH-75 (INSPIRE Faculty Award). MM would like to thank IISER Mohali, India and the DST INSPIRE Faculty award 
INSPIRE-15-0074. MM and MS acknowledge support by the Royal Society International Exchange 2015/R2 program.

\end{document}